\input harvmac

\noblackbox

\let\includefigures=\iftrue
\let\useblackboard=\iftrue
\newfam\black

\includefigures
\message{If you do not have epsf.tex (to include figures),}
\message{change the option at the top of the tex file.}
\input epsf
\def\figin{\epsfcheck\figin}\def\figins{\epsfcheck\figins}
\def\epsfcheck{\ifx\epsfbox\UnDeFiNeD
\message{(NO epsf.tex, FIGURES WILL BE IGNORED)}
\gdef\figin##1{\vskip2in}\gdef\figins##1{\hskip.5in}
\else\message{(FIGURES WILL BE INCLUDED)}%
\gdef\figin##1{##1}\gdef\figins##1{##1}\fi}
\def\DefWarn#1{}
\def\figinsert{\goodbreak\midinsert}
\def\ifig#1#2#3{\DefWarn#1\xdef#1{Fig.~\the\figno}
\writedef{#1\leftbracket Fig.\noexpand~\the\figno}%
\figinsert\figin{\centerline{#3}}\medskip\centerline{\vbox{
\baselineskip12pt\advance\hsize by -1truein
\noindent\footnotefont{\bf Fig.~\the\figno:} #2}}
\bigskip\endinsert\global\advance\figno by1}
\else
\def\ifig#1#2#3{\xdef#1{Fig.~\the\figno}
\writedef{#1\leftbracket Fig.\noexpand~\the\figno}%
\global\advance\figno by1} \fi

\def\doublefig#1#2#3#4{\DefWarn#1\xdef#1{Fig.~\the\figno}
\writedef{#1\leftbracket Fig.\noexpand~\the\figno}%
\figinsert\figin{\centerline{#3\hskip1.0cm#4}}\medskip\centerline{\vbox{
\baselineskip12pt\advance\hsize by -1truein
\noindent\footnotefont{\bf Fig.~\the\figno:} #2}}
\bigskip\endinsert\global\advance\figno by1}

\useblackboard
\message{If you do not have msbm (blackboard bold) fonts,}
\message{change the option at the top of the tex file.}
\font\blackboard=msbm10 scaled \magstep1 \font\blackboards=msbm7
\font\blackboardss=msbm5 \textfont\black=\blackboard
\scriptfont\black=\blackboards \scriptscriptfont\black=\blackboardss

\else

\fi
%

\def\yboxit#1#2{\vbox{\hrule height #1 \hbox{\vrule width #1
\vbox{#2}\vrule width #1 }\hrule height #1 }}
\def\fillbox#1{\hbox to #1{\vbox to #1{\vfil}\hfil}}
\def\ybox{{\lower 1.3pt \yboxit{0.4pt}{\fillbox{8pt}}\hskip-0.2pt}}
%
%


\def\M{{\cal M}}
\def\UM{\tilde{\cal M}}


\def\comments#1{}



\def\II{\relax{I\kern-.10em I}}

\def\IZ{\relax{\rm Z\kern-.34em Z}}
\def\IB{\relax{\rm I\kern-.18em B}}
\def\IC{{\relax\hbox{$\inbar\kern-.3em{\rm C}$}}}
\def\ID{\relax{\rm I\kern-.18em D}}
\def\IE{\relax{\rm I\kern-.18em E}}
\def\IF{\relax{\rm I\kern-.18em F}}
\def\IG{\relax\hbox{$\inbar\kern-.3em{\rm G}$}}
\def\IGa{\relax\hbox{${\rm I}\kern-.18em\Gamma$}}
\def\IH{\relax{\rm I\kern-.18em H}}
\def\II{\relax{\rm I\kern-.18em I}}
\def\IK{\relax{\rm I\kern-.18em K}}
\def\IP{\relax{\rm I\kern-.18em P}}

%

\def\inbar{\,\vrule height1.5ex width.4pt depth0pt}

\def\IR{\relax{\rm I\kern-.18em R}}

\def\simgt{\hskip0.05in\relax{
\raise3.0pt\hbox{ $>$ {\lower5.0pt\hbox{\kern-1.05em $\sim$}} }}
\hskip0.05in}

%


%

\def\lp10{\ell_p^{10}}
\def\lp11{\ell_p^{11}}
\def\R11{R_{11}}

\def\frac#1#2{{#1 \over #2}}



\newdimen\tableauside\tableauside=1.0ex
\newdimen\tableaurule\tableaurule=0.4pt
\newdimen\tableaustep
\def\phantomhrule#1{\hbox{\vbox to0pt{\hrule height\tableaurule width#1\vss}}}
\def\phantomvrule#1{\vbox{\hbox to0pt{\vrule width\tableaurule height#1\hss}}}
\def\sqr{\vbox{%
  \phantomhrule\tableaustep
  \hbox{\phantomvrule\tableaustep\kern\tableaustep\phantomvrule\tableaustep}%
  \hbox{\vbox{\phantomhrule\tableauside}\kern-\tableaurule}}}
\def\squares#1{\hbox{\count0=#1\noindent\loop\sqr
  \advance\count0 by-1 \ifnum\count0>0\repeat}}
\def\tableau#1{\vcenter{\offinterlineskip
  \tableaustep=\tableauside\advance\tableaustep by-\tableaurule
  \kern\normallineskip\hbox
    {\kern\normallineskip\vbox
      {\gettableau#1 0 }%
     \kern\normallineskip\kern\tableaurule}%
  \kern\normallineskip\kern\tableaurule}}
\def\gettableau#1 {\ifnum#1=0\let\next=\null\else
  \squares{#1}\let\next=\gettableau\fi\next}

\tableauside=1.0ex \tableaurule=0.4pt


 %
 %
 \def\eqnn#1{\xdef #1{(\secsym\the\meqno)}\writedef{#1\leftbracket#1}%
 \global\advance\meqno by1\wrlabeL#1}
 \def\eqna#1{\xdef #1##1{\hbox{$(\secsym\the\meqno##1)$}}
 \writedef{#1\numbersign1\leftbracket#1{\numbersign1}}%
 \global\advance\meqno by1\wrlabeL{#1$\{\}$}}
 \def\eqn#1#2{\xdef #1{(\secsym\the\meqno)}\writedef{#1\leftbracket#1}%
 \global\advance\meqno by1$$#2\eqno#1\eqlabeL#1$$}

\global\newcount\itemno \global\itemno=0

\def\itemaut#1{\global\advance\itemno by1\noindent\item{\the\itemno.}#1}


\def\eg{{\it e.g.}}
\def\ie{{\it i.e.}}
\def\cf{{\it c.f.}}

\hyphenation{Di-men-sion-al}



\lref\CornalbaFI{
  L.~Cornalba and M.~S.~Costa,
  ``A new cosmological scenario in string theory,''
  Phys.\ Rev.\ D {\bf 66}, 066001 (2002)
  [arXiv:hep-th/0203031].
}

\lref\CallanEM{
  C.~G.~Callan and F.~Wilczek,
  ``Infrared Behavior at Negative Curvature,''
  Nucl.\ Phys.\ B {\bf 340}, 366 (1990).
}


\lref\HellermanNX{
  S.~Hellerman and I.~Swanson,
  ``Cosmological solutions of supercritical string theory,''
  arXiv:hep-th/0611317.
}

\lref\anyfund{Robert E. Gompf, ``A New Construction of Symplectic
Manifolds," Annals of Mathematics, 2nd Ser., Vol. 142, No. 3. (Nov.,
1995), pp. 527-595.}

\lref\cardy{
  J.~L.~Cardy,
  ``Operator Content Of Two-Dimensional Conformally Invariant Theories,''
  Nucl.\ Phys.\ B {\bf 270}, 186 (1986).
}

\lref\bogomolny{ E.~Bogomolny, ``Quantum and Arithmentical Chaos,"
arXiv:nlin.CD/0312061.}

\lref\joebook{
J.~Polchinski, {\it String Theory}, two vols,
Cambridge (1998).
}

\lref\Link{For a recent discussion, see G. Link, ``Asymptotic
geometry and growth of conjugacy classes of nonpositively curved
manifolds," math.DG/0508346.}

\lref\myers{
  R.~C.~Myers,
 ``New Dimensions For Old Strings,''
  Phys.\ Lett.\ B {\bf 199}, 371 (1987).
}

\lref\othermutation{
  M.~Matone, P.~Pasti, S.~Shadchin and R.~Volpato,
  arXiv:hep-th/0607133;
  S.~Dasgupta and T.~Dasgupta,
  arXiv:hep-th/0609147.
}

\lref\simeonWalcher{
  S.~Hellerman and J.~Walcher,
  ``Worldsheet CFTs for flat monodrofolds,''
  arXiv:hep-th/0604191.
}

\lref\margulis{G.A. Margulis, ``Applications of Ergodic Theory to
the Investigation of Manifolds of Negative Curvature", Funct. Anal.
Appl. 3 (1969), 335-336.}

\lref\micha{
  N.~A.~Nekrasov,
  ``Milne universe, tachyons, and quantum group,''
  Surveys High Energ.\ Phys.\  {\bf 17}, 115 (2002)
  [arXiv:hep-th/0203112];
  B.~Pioline and M.~Berkooz,
  ``Strings in an electric field, and the Milne universe,''
  JCAP {\bf 0311}, 007 (2003)
  [arXiv:hep-th/0307280];
  M.~Berkooz, B.~Pioline and M.~Rozali,
  ``Closed strings in Misner space,''
  JCAP {\bf 0408}, 004 (2004)
  [arXiv:hep-th/0405126].
}

\lref\FS{
  W.~Fischler and L.~Susskind,
  ``Dilaton Tadpoles, String Condensates And Scale Invariance,''
  Phys.\ Lett.\ B {\bf 171}, 383 (1986).
}

\lref\ofer{O. Aharony and E. Silverstein, ``Supercritical Stability,
Transitions and (Pseudo)tachyons," hep-th/0612031.}

\lref\Riccisolns{J. Isenberg and M. Jackson, ``Ricci flow of Locally
Homogeneous Geometries on Closed Manifolds," J. Di. Geom. 35, p.
723-741 (1992); D. Knopf and K. McLeod, ``Quasi-Convergence of Model
Geometries under the Ricci flow," Comm. Anal. Geom. 9, p. 879-919
(2001); J. Lott, ``On the Long-time behavior of type-III Ricci flow
solutions," math.DG/0509639.}

\lref\constraints{
  R.~Bartnik and J.~Isenberg,
  ``The constraint equations,''
  arXiv:gr-qc/0405092.
}

\lref\Dproject{D. Green, A. Lawrence, J. McGreevy, D. Morrison, E.
Silverstein, ``Dimensional Duality," in progress.}

\lref\topcosmo{ Testing Global Isotropy of Three-Year Wilkinson
Microwave Anisotropy Probe (WMAP) Data: Temperature Analysis
Authors: Amir Hajian (PRINCETON), Tarun Souradeep (IUCAA)
astro-ph/0607153; Creation of a compact topologically nontrivial
inflationary universe. Andrei Linde (Stanford U., Phys. Dept.) . Aug
2004. 8pp. Published in JCAP 0410:004,2004 e-Print Archive:
hep-th/0408164}

\lref\newhandle{
  A.~Saltman and E.~Silverstein,
  ``A new handle on de Sitter compactifications,''
  JHEP {\bf 0601}, 139 (2006)
  [arXiv:hep-th/0411271].
}

\lref\tendim{
  J.~A.~Harvey, S.~Kachru, G.~W.~Moore and E.~Silverstein,
  ``Tension is dimension,''
  JHEP {\bf 0003}, 001 (2000)
  [arXiv:hep-th/9909072].
}

\lref\BV{
  N.~L.~Balazs and A.~Voros,
  ``Chaos on the pseudosphere,''
  Phys.\ Rept.\  {\bf 143}, 109 (1986).
}

\lref\delaharpe{P.~de la Harpe, ``Topics in Geometric Group Theory,"
U. Chicago Press (2000).}

\lref\webquestion{{\tt http://zebra.sci.ccny.cuny.edu/web/nygtc/problems/probgrowth.html}}

\lref\grigreview{Rostislav Grigorchuk, Igor Pak, ``Groups of
Intermediate Growth: an Introduction for Beginners," [arXiv:
math.GR/0607384] }

\lref\tomasiello{discussions with A. Tomasiello}

\lref\milnor{J. Milnor, ``A note on curvature and fundamental group,"
J. Diff. Geom, 1968 p. 1-7.}

\lref\manning{A. Manning, ``Relating Exponential Growth in a
Manifold and its Fundamental Group",
Proceedings of the AMS, {\bf 133} Number 4, 995.}

\lref\manningtop{A. Manning, ``Topological Entropy for Geodesic
Flows," Annals of Mathematics, 2nd Ser., Vol. 110, No. 3. (Nov.,
1979), pp. 567-573. }

\lref\wittenAdS{
  E.~Witten,
  ``Anti-de Sitter space and holography,''
  Adv.\ Theor.\ Math.\ Phys.\  {\bf 2}, 253 (1998)
  [arXiv:hep-th/9802150].
}

\lref\tasilectures{
  E.~Silverstein,
  ``Tasi / Pitp / Iss Lectures On Moduli And Microphysics,''
  arXiv:hep-th/0405068.
}

\lref\mss{
  E.~Silverstein,
  ``(A)dS backgrounds from asymmetric orientifolds,''
  contribution to Strings 2001 [hep-th/0106209];
  A.~Maloney, E.~Silverstein and A.~Strominger,
  ``De Sitter space in noncritical string theory,''
  in {\it Cambridge 2002: The future of theoretical physics and
  cosmology}, 570-591
 [hep-th/0205316].}

\lref\kklt{  
  S.~Kachru, R.~Kallosh, A.~Linde and S.~P.~Trivedi,
  ``De Sitter Vacua In String Theory,''
  Phys.\ Rev.\ D {\bf 68}, 046005 (2003)
  [hep-th/0301240].
}

\lref\mutation{
  E.~Silverstein,
  ``Dimensional mutation and spacelike singularities,''
  Phys.\ Rev.\ D {\bf 73}, 086004 (2006)
  [arXiv:hep-th/0510044].
}

\lref\kutsei{
  D.~Kutasov and N.~Seiberg,
   ``Number Of Degrees Of Freedom, Density Of States And Tachyons In String
  Theory And CFT,''
  Nucl.\ Phys.\ B {\bf 358}, 600 (1991).
}

\lref\lms{
  H.~Liu, G.~W.~Moore and N.~Seiberg,
``Strings in a time-dependent orbifold,''
  JHEP {\bf 0206}, 045 (2002)
  [arXiv:hep-th/0204168].
}

\lref\threesum{
  M.~Anderson, S.~Carlip, J.~G.~Ratcliffe, S.~Surya and S.~T.~Tschantz,
  ``Peaks in the Hartle-Hawking wave function from sums over topologies,''
  Class.\ Quant.\ Grav.\  {\bf 21}, 729 (2004)
  [arXiv:gr-qc/0310002];
  M. Freedman, semiprivate communication
  ({\tt http://math.ucsb.edu/~drm/GTPseminar/}).
}

\Title{\vbox{\baselineskip12pt\hbox{hep-th/0612121}
\hbox{NSF-KITP-06-127}\hbox{SU-ITP-06/33}\hbox{SLAC-PUB-12250}\hbox{MIT-CTP/3795}}}
{\vbox{ \centerline{} \centerline{New Dimensions for Wound Strings:}
\medskip\centerline{
{\authorfont The Modular Transformation of Geometry to Topology}}
%
%
%
%
%
%
%
%
%
%
}}
\bigskip
\bigskip
\centerline{John McGreevy$^1$, Eva Silverstein$^{2,3}$,
and David Starr$^{2,3}$}
\bigskip
\centerline{$^1${\it Center for Theoretical Physics,
Massachusetts Institute of Technology, Cambridge, MA 02139}}
\bigskip
\centerline{$^2${\it SLAC and Department of Physics, Stanford
University, Stanford, CA 94305-4060}}
\bigskip
\centerline{$^3${\it Kavli Institute for Theoretical Physics,
University of California, Santa Barbara, CA 93106-4030}}
\bigskip
\bigskip
\noindent

We show, using a theorem of Milnor and Margulis, that string theory
on compact negatively curved spaces grows new effective dimensions
as the space shrinks, generalizing and contextualizing the results
in \mutation. Milnor's theorem relates negative sectional curvature
on a compact Riemannian manifold to exponential growth of its
fundamental group, which translates in string theory to a higher
effective central charge arising from winding strings. This
exponential density of winding modes is related by modular
invariance to the infrared small perturbation spectrum. Using
self-consistent approximations valid at large radius, we analyze
this correspondence explicitly in a broad set of time-dependent
solutions, finding precise agreement between the effective central
charge and the corresponding infrared small perturbation spectrum.
This indicates a basic relation between geometry, topology, and
dimensionality in string theory.

\bigskip
\Date{December 2006}



\newsec{Introduction}

Negatively curved spaces constitute a basic class of geometries, and
hence a basic class of potential target spaces for string theory. At
large curvature radius, one has an easily controlled description of
the system via general relativity; in vacuum, for example, such
spaces expand slowly with time in a way described by self-consistent
solutions of the equations of motion.

At the same time, perturbative string theory probes spacetime geometry
and topology differently from point particle theory.  For example, on
a space with nontrivial fundamental group the perturbative string
spectrum contains winding string sectors; again this is a
well-controlled statement at large radius.

A fundamental relation between the geometry and the topology of
compact negatively curved spaces is described by Milnor's theorem
\milnor. This says that negative sectional curvature and compactness
imply {\it exponential growth} of the fundamental group
\refs{\milnor,\margulis}. Exponential growth of the fundamental
group implies \margulis\ that the number density $\rho(l)$ of
independent geodesics of length $l$ is an exponentially growing
function of $l$:
\eqn\growthrho{\rho(l) \sim l^\alpha e^{l/l_0}}
for some finite constants $l_0$ and $\alpha$.

In the context of perturbative string theory on such a space, in the
large radius regime, this correspondence has a simple physical
interpretation (introduced in a special case in \mutation). The
exponential growth in the fundamental group implies an exponential
growth in the density of states of string winding modes (independent
of oscillator modes), exhibiting compact negatively curved target
spaces as supercritical backgrounds of string theory.

Moreover, the exponential growth in the density of states is related
by a modular transformation to the IR behavior of the partition
function \refs{\cardy, \kutsei}.  The spectrum of IR perturbations
is in turn sensitive to the curvature and
compactness of the spatial slices.

In this work, we analyze in some detail the physical realization of
this correspondence in a broad class of background solutions,
exhibiting precise agreement with this prediction of modular
invariance.  Although we will not compute the full modular-invariant
partition function, we will explicitly work out the leading effects
in the deep IR and deep UV parts of the spectrum, which depend in a
clear way on the compactness, curvature, and topology of our target
space. In these limits, the semiclassical approximation controlling
both the infrared spectrum and the winding strings provides a
straightforward means for consistently analyzing the one-loop
partition function in a time-dependent background.\foot{See, \eg,\
\refs{\CornalbaFI,\micha, \lms} for previous works computing and
interpreting a full partition function in a Lorentzian-signature
target space. There remain significant subleties with these models
partly to do with their construction as a global orbifold of
Minkowski space which introduces extra challenges in the very early
time physics. In the present work we will evade these subtleties by
working in a regime where IR quantum field theory and semiclassical
cosmic strings are easily controlled, and by avoiding a global
orbifold construction.} One interesting outcome is that this check
of modular invariance works timeslice by timeslice, \ie,\ it works
before performing the path integration over the time variable.


It is interesting also to ask what the effective dimensionality
becomes at small radius. In the case of genus $h$ Riemann surfaces,
numerous diagnostics -- including a variant of Buscher's worldsheet
path integral argument for T-duality, D-brane systems, and
algebro-geometric genericity arguments -- all suggest that the
system can naturally cross over to a supercritical theory on the
$2h$-dimensional Jacobian torus, with a rolling tachyon producing a
transition from this to the large radius expanding space
\Dproject.\foot{More generally, for higher dimensional examples the
Albanese variety arises at small radius.}

Altogether, we are being led to a generalization of T-duality and
the Calabi-Yau/Landau-Ginzburg correspondence applicable to the
generic case of curved compact target spaces. This provides a new
mechanism by which spatial dimensions emerge, arising in a novel way
from nontrivial topology.

The paper is organized as follows.  We start by reviewing the
modular invariance relation between the UV and IR limits of the
partition function.  Next we analyze these limits for solutions of
the equations of motion with compact spatial slices of constant
negative curvature, using their realization as freely acting
orbifolds of hyperbolic $n$-dimensional space.  We find precise
agreement with the modular invariance prediction for the relation
between the deep UV and deep IR contributions to the partition
function. We consider a more subtle case of exponential growth,
obtained from compact quotients of the Sol geometry, and a subtle
case of power law growth, obtained from compact quotients of the Nil
geometry.  In all these homogeneous cases, modular invariance and
the growth of effective dimensions checks out explicitly.  We then
discuss more general cases, analogies to other mechanisms for
growing dimensions in string theory, and applications to
mathematical and physical questions in the last two sections.

\newsec{Structure of the One-loop Amplitude}

Let us use the conventions of Polchinski's book \joebook. The one-loop vacuum
amplitude in perturbative closed string theory takes the form
\eqn\genform{\int {d^2\tau \over{\tau_2^2}} Z_1(\tau)={\rm Tr} \int
{d^2\tau\over{4\tau_2}} (-1)^F q^{L_0}\bar q^{\tilde L_0},}
where $\tau$ is the modular parameter of the worldsheet torus, $F$
is the spacetime fermion number, $q=e^{2\pi i\tau}$, and $L_0\pm
\tilde L_0$ are the worldsheet Hamiltonian and momentum generators.
Here $Z_1(\tau)$ is defined so as to be modular invariant:
$Z_1(\tau) = Z_1 ( \tau+1) = Z_1(-1/ \tau)$.

The $\tau_2\to \infty$ region of the integral is dominated by the
deep IR part of the spectrum; $Z_1$ can be expanded as
\eqn\IRZ{Z_1/\tau_2\to q^{L_{0,\rm{min}}}\bar q^{\tilde L_{0,\rm{min}}}.}

Let us first review the IR and UV behavior of $Z_1$ in the
10-dimensional classical backgrounds of the superstring (including
the tachyonic cases) and the 26-dimensional unstable background of
the bosonic string. In terms of the spacetime masses and momenta,
for level-matched contributions we have
\eqn\massL{L_0 = (k^2+m^2)\alpha'/4 = \tilde L_0, }
so the IR limit of the partition function has an integrand going
like\foot{Here, as will be our custom, we ignore the subleading
polynomial dependence of the asymptotic form.  Equivalently, our
statements that two functions are asymptotic should be interpreted as
statements about the corresponding logarithms.}
\eqn\ZIR{Z_{\rm{IR}}
\equiv Z_1|_{\tau_2\to\infty}\sim e^{-\pi \tau_2\alpha'm^2_{\rm{min}}}.}

The $\tau_2\to 0$ region is dominated by the deep UV.
We can compute directly the UV asymptotics of the
torus amplitude using the density of states for large
masses,
\eqn\densm{\rho(m)\sim e^{m\sqrt{\alpha'}\pi\sqrt{2c_{\rm{eff}}/3}} , }
where $c_{\rm{eff}}$ is the effective central charge.
Summing over $m$ with this measure
by saddle
point approximation gives
\eqn\ZUVc{Z_{\rm{UV}} \equiv Z_1|_{\tau_2 \to 0} \sim \sum_m
\rho(m)e^{-\pi \tau_2\alpha'm^2} \sim e^{{{\pi
c_{\rm{eff}}}\over{6\tau_2 }}}.}
By modular
invariance, the dominant contribution in the $\tau_2 \to 0$ region must go like
\eqn\ZUV{Z_{\rm{UV}}\sim e^{-\pi \alpha'm^2_{\rm{min}}/\tau_2}.}
So we have
\eqn\creln{c_{\rm{eff}}=-6\alpha' m^2_{\rm{min}}.}
As a check, note that in the case of the type 0 theory, the lightest
state is a tachyon $\alpha' m^2_{\rm{min}}=-2$.  This corresponds to
$c_{\rm{eff}}=12$, the correct value once one subtracts the 2 dimensions
cancelled out by the ghosts.  Similarly in the bosonic string, the
lightest state is a tachyon with $\alpha'm^2_{\rm{min}}=-4$,
corresponding to $c_{\rm{eff}}=24$.

For models in which the mass squared is nonnegative, such as
spacetime-supersymmetric strings, there is no such exponential IR
divergence in the partition function. Hence, by modular invariance,
there is also no surviving contribution of order
$e^{\rm{const}/\tau_2}$. This requires a dramatic cancellation going
well beyond the necessary condition that the effective central charges
of the spacetime bosons and fermions agree,
$c_{\rm{eff}}^F=c_{\rm{eff}}^B$ \kutsei.

Theories which do have such IR divergences might naively be expected to
universally suffer from dramatic instabilities on equal footing with
those in the bosonic string theory.  However, this is not the case;
IR divergences can arise from modes, {\it pseudotachyons}, whose
condensation does not cause a large back reaction (as occurs in the
standard case of inflationary perturbation theory, as well as in
more formal time-dependent string backgrounds
\refs{\HellermanNX,\ofer}). This is the case we will encounter in
our examples. In the next section we will study the UV and IR limits
of the partition function, and the modular transformation between
them, in the case of spacetimes with compact negatively curved
spatial slices.

\newsec{Exponential Growth:  Examples}

Any compact manifold of negative sectional curvature has a
fundamental group of exponential growth \milnor, meaning that the
number of independent elements grows exponentially with minimal word
length in any basis of generators of the group. It follows
\margulis\ that the number of free homotopy classes also grows
exponentially with geodesic length.

In this section, we reproduce this result in constant curvature
examples using modular invariance in string theory.  This analysis
will provide precise results for the IR and UV limits of the
partition function, and their modular equivalence, exhibiting these
compact negatively curved target spaces as supercritical backgrounds
of string theory as suggested in \mutation.  We will then consider
more general examples and applications.

\subsec{Orbifolds of Hyperboloids}

Consider first $n$-dimensional manifolds of constant negative
curvature. These have a universal cover $\IH_n$, the $n$-dimensional
noncompact hyperbolic space with metric
\eqn\metcov{ds^2_{\IH_n}=R_0^2 \,(dy^2+\sinh^2y\, d\Omega^2) .}
For some of this discussion, we will set $R_0=1$ for simplicity. For
$n\le 9$, a solution of the equations of motion of critical
superstring theory is obtained by considering spacial slices
\metcov\ expanding in time, combined with a transverse
$(9-n)$-dimensional
Ricci flat space $X_\perp$ which we will take to be
compact with metric $ds^2_\perp$:
\eqn\tmetcov{ds^2 =-dt^2+({t/t_0})^2 ds^2_{\IH_n}+ds^2_\perp .}
This, of course, is nothing but the inside of a light cone in flat
spacetime $M_{n,1}$ (times $X_\perp$), with time coordinate chosen
so that the $M_{n,1}$ component has spatial slices of constant
negative curvature.  The constant $t_0$ will often be set to $t_0 =
R_0$.

Many other solutions are obtained by orbifolding the space $\IH_n$ by a
freely acting discrete group of isometries, $\Gamma$, such that the
resulting space ${\cal M}_n\equiv \IH_n/\Gamma$ is compact.  This
space has $\pi_1({\cal M}_n)\cong\Gamma$ and metric
\eqn\tmetM{ds^2 =-dt^2+({t/t_0})^2 ds^2_{{\cal
M}_n=\IH_n/\Gamma}+ds^2_\perp,}
which is locally the same as flat spacetime \tmetcov.  Globally, the
spacetime \tmetM\ is very different from the noncompact flat covering
spacetime \tmetcov; it is not Poincar\'e invariant, it is spatially
compact, and it has a nontrivial (and exponentially growing)
fundamental group.  As we will see, the latter two features affect the
IR and UV behavior of the one-loop partition function in a way
consistent with modular invariance.

We will consider string theory on these target spaces and compare
and contrast the structure of the one-loop partition function in the
flat spacetime \tmetcov\ to that in the expanding compact daughter
space \tmetM. Note that the model we analyze is not necessarily the
same as any global orbifold of Minkowski space by $\Gamma$.  We do
find it useful to obtain our compact {\it spatial} slices by
projection from the $\IH_n$ spatial slices of \tmetcov, which sit
within the future light cone of a point in Minkowski space. In any
case we will restrict attention to the late-time large-radius system
and require of the initial state only that it lead to a smooth
large-radius geometry at late times.

Before proceeding, let us explain the framework we will use to
analyze the perturbative string theory. In general, time-dependent
target spaces of string theory are difficult to formulate
explicitly. However, solving Einstein's equations provides a
semiclassical solution to the conditions for worldsheet conformal
invariance, valid as the leading approximation in a self-consistent
expansion at large curvature radius.

Formally, the worldsheet path integration then generates appropriate
corrections to the semiclassical action obtained from this background
solution; one integrates over the conformal factor in the metric,
leading to a system whose total central charge is zero.  However, a
priori, the worldsheet path integral is not manifestly convergent. The
path integral may be rendered convergent by a Wick rotation in some
circumstances (such as in flat spacetime), but most backgrounds do not
have a smooth Euclidean continuation. In quantum field theory on
curved spacetime, one can work directly in Lorentzian signature,
including a convergence factor in the calculation of amplitudes.  A
similar method may apply to the full worldsheet string path integral;
indeed some works have succeeded in calculating the one-loop partition
function of some simple time-dependent backgrounds using a Euclidean
worldsheet and Lorentzian target space \micha. For these reasons, we
expect our backgrounds to provide consistent solutions, which at late
times are accurately described by perturbative string theory and its
general relativistic approximation.  Our computations can be viewed as
a test of this proposition, more specifically as a concrete test of
modular invariance of the putative worldsheet string path integral,
and of the existence of initial states consistent with our
large-radius phase.

We will begin with a discussion of the bosonic contributions to the
one-loop partition function in the IR and the UV.  In \S3.4\ we will
comment on the fermionic contributions, arguing both in the IR and
the UV that they do not cancel out the effect.

\subsec{IR Limit}

Consider a massless scalar field $\eta(t,y,\Omega)$, a classical
modulus of $X_\perp$, propagating on $M_{n,1}$, coordinatized as in
\tmetcov.\foot{ A similar but more indexful analysis applies to
perturbations of the metric.}  We will be interested in the spectrum
of modes which propagate in the one-loop partition function.

The following feature of this spectrum will play an important role.
In the parent theory \tmetcov, a complete set of normalizable
eigenfunctions of the Laplacian $\nabla^2_{\IH_n}$ on the spatial
slices $\IH_n$ is gapped \refs{\CallanEM,\BV}, due to the rapid
growth of the volume as a function of proper distance $y$ at large
$y$; this can be shown as follows.

Consider the eigenvalue equation
\eqn\lapH{\nabla^2_{\IH_n}\tilde f_k(y,\Omega) = -k^2 \tilde
f_k(y,\Omega),}
where $\nabla^2_{{\IH}_n}$ is the spatial Laplacian:
\eqn\spaceLap{\nabla^2_{{\IH}_n}
= \frac{1}{\sinh^{n-1}y}
  \partial_y(\sinh^{n-1}y\, \partial_y )
  + \frac{1}{\sinh^2 y}\nabla^2_{\Omega}.}
Separate variables as $\tilde f_k(y,\Omega)=f_{k,L}(y)Y_L(\Omega)$ with
$Y_L$ denoting spherical harmonics that satisfy $\nabla^2_\Omega
Y_L=-L^2Y_L$. Then \lapH\ reduces to
\eqn\lapHy{{1\over \sinh^{n-1}y}\del_y(\sinh^{n-1}y\,\del_y f_{k,L})-{L^2\over
\sinh^2 y}f_{k,L} = -k^2 f_{k,L}.}
At large $y$ this becomes
\eqn\lapHlargey{e^{-(n-1)y}\del_y\left(e^{(n-1)y}\del_y f_{k,L}\right) = -k^2 f_{k,L} ,}
so that $f_{k,L}$ is in fact independent of $L$ at large $y$.  We
can solve this with the ansatz $f_{k,L}(y\to\infty) \sim f_0
e^{\lambda y}$ which, when plugged into \lapHlargey, yields the two
independent solutions $\lambda_\pm = {1\over
2}(-(n-1)\pm\sqrt{(n-1)^2-4k^2})$.   Consider the range $k^2<
(n-1)^2/4$.  In this range, the solution smooth in the interior is
the solution with $\lambda=\lambda_+$, as noted in
\wittenAdS.\foot{This is straightforward to establish by rewriting
the equation \lapHy\ as an equation for $\hat f=e^{(n-1)y/2}f$ and
using it to constrain the relative signs of $\hat f$ and its
derivatives.  This enables one to rule out the possibility of a
solution matching the $\lambda=\lambda_-$ behavior at infinity with
the nonsingular (constant) solution at $y=0$.}

However, in this range $k^2< (n-1)^2/4$ the solution is not
normalizable:
\eqn\fnorm{\int^\infty dy \sqrt{-G} f_{k,L}^*(y)f_{k,L}(y)\sim \int^\infty
dy \,e^{(n-1)y} |f_0|^2 e^{\left(-(n-1)+\sqrt{(n-1)^2-4k^2}\right)y} = \infty.}
That is, there is a gap in the spectrum:
\eqn\gap{k^2 \geq k_{\rm{gap}}^2 = {(n-1)^2\over 4}  .}

The time dependence in \tmetcov\ compensates for this, lowering the
effective mass squared of each mode expanded about the
time-dependent background, similarly to the situation in the
supercritical linear dilaton theory \myers.   To see this, consider
massless on-shell modes of $\eta$ in the background \tmetcov.  The
equation of motion for $\eta$ is
\eqn\lapeq{\nabla^2 \eta = 0,}
where $\nabla^2$ is the Laplacian acting on the full spacetime:
\eqn\Laptot{\nabla^2
= - \frac{1}{t^n}\partial_t(t^n\partial_t)
  + \frac{1}{t^2}\nabla^2_{\IH_n}.}
We can decompose $\eta$ into modes as $\eta(t, y,\Omega) =
u(t)f_{k,L}(y)Y_L(\Omega)$, with $f_{k,L}$ and $Y_L$ as above.  Then the
equation of motion \lapeq\ for $\eta(t,y,\Omega)$ reduces to an equation for
$u(t)$ alone:
\eqn\onshell{-{1\over t^n}\del_t(t^n\del_t u(t))= {k^2\over
t^2}u(t).}
This has solutions
\eqn\usol{u(t)\propto t^\alpha, \qquad \alpha={1\over
2}\left(-(n-1)\pm\sqrt{(n-1)^2-4k^2}\right).}
So we see that precisely for $k$ above the gap, satisfying \gap, the
massless solutions oscillate in time.  Moreover, the frequency goes to
zero as $k^2\to (n-1)^2/4$, so the modes behave like massless
particles with zero momentum for $k=k_{\rm{gap}}$.  The effects of the
spatial curvature and the effects of the time-dependent evolution
precisely cancel.

This effect can be seen in the propagator constructed in the low
energy field theory, which is closely related to the IR limit of the
one-loop amplitude of interest in the perturbative string theory.
This is obtained by solving
\eqn\LapD{\nabla^2 D(x,x')=-\delta(x-x')/\sqrt{-g}}
in terms of a complete basis of functions.

It proves convenient to consider the complete set of functions
\eqn\sepvar{\psi_{\omega,k,L}(t,y,\Omega) = u_\omega(t)f_{k,L}(y)Y_L(\Omega),}
with $f_{k,L}(y)$ and $Y_L(\Omega)$ as described above and with
$u_\omega(t)$ given by
\eqn\uform{ u_\omega(t) = t^{\frac{1-n}{2}}e^{i\omega \ln t}.}
These satisfy
\eqn\ueq{ -{1\over t^n}\del_t(t^n\del_t u_\omega)
= \frac{1}{t^2}\left(\omega^2+\frac{(n-1)^2}{4}\right)u_\omega.}

The full spacetime Laplacian \Laptot\ acting on $\psi_{\omega,k,L}$
will then give
\eqn\nonebasis{\nabla^2
\psi_{\omega,k,L} =
\frac{1}{t^2}\left(\omega^2-k^2+\frac{(n-1)^2}{4}\right)
\psi_{\omega,k,L}.}
In \nonebasis\ the gap is evident.  At large $y$ the modes \sepvar\
take the form
\eqn\largeymodes{\psi_{\omega,k,L}(t,y,\Omega) \sim t^{\frac{1-n}{2}}e^{i\omega
\ln t} e^{\left(\frac{1-n}{2}\pm i\sqrt{k^2-k_{\rm{gap}}^2}\right)y}\,Y_L(\Omega).}

The Laplacian \Laptot\ is the low-energy approximation to the
worldsheet Hamiltonian ${\cal H}\equiv L_0+\tilde L_0$.  Note that
we are not working in a basis of eigenfunctions of ${\cal H}$, but
instead are considering a basis of eigenfunctions of $t^2{\cal H}$.
The one-loop vacuum amplitude involves a trace \genform, which can be
evaluated in any basis.  The basis \uform\ is slightly more
convenient for our calculations than the basis of eigenfunctions of
${\cal H}$ simply because the latter are more complicated Bessel
functions.

The propagator can be expressed as
\eqn\prop{D(x,x')=(tt')^{{1-n}\over 2}\int{d\omega\over{2\pi}}
e^{i\omega(\ln t-\ln t')}
\sum_{L}
\int_{k^2\ge (n-1)^2/4}
{
f_{L,k}^*(y)f_{L,k}(y')Y_L^*(\Omega)Y_L(\Omega')
\over -\omega^2+k^2-(n-1)^2/4}.}
It is straightforward to check that this provides a solution to the
equation \LapD\ for the two-point Green's function.

In general, physical amplitudes depend on a choice of vacuum state,
and as mentioned above we assume initial conditions consistent with
a late-time phase with weakly curved spatial slices.
In the Friedmann equation with negative spatial curvature,
radiation, and matter, $(\dot
a/a)^2=1/a^2+g_s^2[(\rho_r/a^4)+(\rho_m/a^3)]$, curvature dominates
at late times. The presence of a bath of particles, which is the
typical relic of working in another vacuum (distinct from that
killed by the annihilation operators corresponding to our modes
\sepvar), will not significantly alter our results.

The one-loop vacuum amplitude in the point-particle approximation is
%
\eqn\vacampl{
\eqalign{{\rm Tr ~log}({\cal H}) &= \int d^dx\sqrt{-g}\,\Lambda(x)   \cr
&= \int dtdyd\Omega\, t^n \sinh^{n-1}y
\, \int{d\tilde\omega\over{2\pi}}
\sum_L
\int_{k^2 \geq (n-1)^2/4}
\left| f_{L,k}(y)Y_L(\Omega)\right|^2
\cr & \qquad\qquad\times
\log\biggl({\tilde\omega^2+k^2-{(n-1)^2/ 4}\over t^2}
\biggr), \cr }}
where we Wick rotated the $\omega$ integral via
$\omega=i\tilde\omega$.

This can be expressed in a Schwinger formalism appropriate to the IR
limit of the first-quantized string path integral as
\eqn\Schwing{\eqalign{& \int dtdyd\Omega \,t^n \sinh^{n-1}y\,
\int{d\tilde\omega\over{2\pi}} \sum_{L} \int_{k^2\ge (n-1)^2/4}
\left| f_{L,k}(y)Y_L(\Omega)\right|^2
\cr &
~~~~~~~~~ \times \int {d\tau_2\over
\tau_2}\exp{\left( -\pi\alpha'\tau_2\biggl({\tilde\omega^2+k^2-(n-1)^2/4\over t^2}
\biggr)  \right).}\cr }}
The net effect agrees with the standard flat space result. The
fluctuations of $\eta$ are massless, and the one-loop partition
function has no IR divergence as $\tau_2\to\infty$, because
normalizability requires $k^2\ge (n-1)^2/4$.

In the compact daughter space ${\cal M}_n=\IH_n/\Gamma$, in
contrast, the constraint of normalizability on the spatial slices
${\cal M}_n=\IH_n/\Gamma$ is much less restrictive: since the space
is compact, the zero mode is normalizable, and there is no gap in
the spectrum.  Hence on ${\cal M}_n$, one sums over $k$ values
including $k=0$, leading to an IR-divergent term from the $\eta$
contribution to the one-loop amplitude:
\eqn\Zorb{Z_{\rm{IR}}\sim
e^{\pi\alpha'\tau_2(n-1)^2/(4t^2)}.}
The time dependence still lowers the effective mass squared of the
mode. The magnitude of the negative effective mass squared of the
zero mode is given precisely by the gap in the spectrum on the
parent space $\IH_n$.

That this mode is only {\it pseudo}tachyonic rather than being a
catastrophic instability can be seen as follows. The action for the zero mode
is of the form $S_0=\int dt \,t^n(\dot\eta)^2$, with solutions
\eqn\soleta{\eta=\eta_0+\eta_1/t^{n-1}.}
The resulting energy density is of order $\dot\eta^2\sim 1/t^{2n}$,
which is much smaller than Hubble squared $H^2\sim 1/t^2$ for our
original vacuum solution. Hence the back reaction of the rolling
$\eta$ mode on the spacetime geometry is negligible. Moreover, the
solutions \soleta\ show that the Hubble friction from the
expanding space ${\cal M}_n$ prevents the modulus field $\eta$ from
straying far from its initial value.

More generally, for massive or tachyonic particles in the parent
theory (with nonzero mass squared $m^2$), we can carry out the above
analysis working with a basis of mode solutions satisfying (\cf\
\nonebasis)
\eqn\nonebasismass{\left(\nabla^2 - m^2\right) \psi_{\omega,k,L} =
\left[\frac{1}{t^2}\left(\omega^2-k^2+\frac{(n-1)^2}{4}\right)-m^2\right]
\psi_{\omega,k,L}.}
In any perturbative string theory, the minimal mass squared obtained
in the daughter theory is that in the parent theory $m^{(0)\
2}_{\rm{min}}$, offset by the gap $-(n-1)^2/(4 t^2)$.

For example, we can formally apply our methods to the bosonic string
theory, whose lowest lying mode in the flat spacetime \tmetcov\ is a
tachyon of mass squared $m^2_T=-4/\alpha^\prime$.  Once we project to
the daughter space ${\cal M}_n = \IH_n / \Gamma$, the minimal mass
squared in the bosonic theory drops to
\eqn\minB{m^{(B)\ 2}_{\rm{min}}(t)=-{4\over\alpha^\prime}-
{(n-1)^2\over{4t^2}}.}
%


\subsec{Modular Transform and UV Limit}

In the previous subsection, we found a gap in the spectrum of
spatial modes at
%
%
\eqn\kgap{ {k_{\rm{gap}}^2\over t^2}={(n-1)^2\over {4 t^2}}.}
This led to a minimal mass squared
\eqn\gapmass{ m^2_{\rm{min}}=-{(n-1)^2\over {4 t^2}}+m^{(0)\
2}_{\rm{min}},}
where $m^{(0)\ 2}_{\rm{min}}$ is the minimal mass squared in
Minkowski space (which is zero in the superstring, and equal to
$m_T^2=-4/\alpha^\prime$ in the bosonic string). We obtained this by
evaluating the partition function in the deep IR limit, where
low-energy effective field theory applies. Now we would like to
evaluate the partition function in the deep UV limit, as in \ZUVc,
to check modular invariance.

Before implementing this, we must address one subtlety. In general,
the string partition function in a consistent background is only
guaranteed to be conformally invariant (and hence modular invariant)
after performing the path integral over all the embedding
coordinates, including $t$; in the Hamiltonian formalism,
equivalently, modular invariance only holds in general after taking
the trace in \genform. Using our complete basis of functions
\sepvar\nonebasis, we will find that our test of modular invariance
works before performing the $t$ integral.  A priori, this is not
necessary, but it is a sufficient condition for passing this test of
modular invariance.\foot{The fact that it works for every $t$ may
follow from a more extensive analysis of one-loop non-vacuum
S-matrix amplitudes involving a set of different scattering
processes each of which preserve modular invariance but which
localize interactions at different times.  It is somewhat
reminiscent of ``fiber-wise" string dualities.}

Let us now turn to the test of modular invariance.  Plugging
our result for $m_{\rm{min}}$ \gapmass\ into
the modular relation \ZUV\ gives
the small-$\tau_2$ behavior demanded by modular invariance:
\eqn\ZUVorb{Z_{\rm{UV}}\sim {\rm exp}\left({{\pi\alpha' (n-1)^2}\over{4
t^2\tau_2}}+{{\pi c_{\rm eff}^{\rm{crit}}}\over{6\tau_2}}\right).}
where $c_{\rm eff}^{\rm{crit}}$ is the value of the effective central
charge in the critical theory on Minkowski space (reviewed in \S2\
above). Comparing to \ZUVc\ yields
\eqn\corb{c_{\rm{eff}}-c_{\rm
eff}^{\rm{crit}}={3\alpha^\prime(n-1)^2\over 2 t^2} .}
This means that modular invariance can be satisfied in the late-time
limit only if the compact negatively curved manifold ${\cal M}_n$ provides
a new contribution to the exponential density of states.  In the
superstring this further requires the new contribution to survive the
sum over bosonic and fermionic contributions, an issue we will discuss
below in \S3.4.

The projection of $\IH_n$ by freely acting isometries $\Gamma$ has
several effects, which modify both the IR and UV limits of the
spectrum. First, it compactifies the space, so that new modes become
normalizable; this introduces a new state into the IR spectrum which
causes a divergence in the one-loop partition function as we saw in
\S3.2. Second, it restricts unwound modes to invariant states.
Third, it introduces new sectors of winding strings; we will see
that these winding strings contribute a Hagedorn density of states
to the UV limit of the spectrum in just the right way to provide the
requisite effective central charge \corb.\foot{In our case the
winding strings which contribute to the exponential density are in
one-to-one correspondence with free homotopy classes (conjugacy
classes of the fundamental group).  In more general examples, one
might find additional contributions from metastable closed geodesics
that are not supported by topology. This does not happen in cases
where the sectional curvature is bounded above by a negative value
\Link, in particular in our cases of constant curvature.}


The partition function includes a sum over winding sectors and a sum
over states arising from transverse motion of the winding strings,
as well as a sum over eigenstates of $L_0,\tilde L_0$ in each
winding sector (which we will refer to as ``oscillator modes"). The
mass of each winding mode is
\eqn\windmass{m_w^2=R^2/\alpha^{\prime 2}}
when the string winds a cycle of length $l\equiv 2\pi R$.

We are studying the one-loop partition function in the limit where it
degenerates into a long thin tube.  In the previous section, we
studied this in the channel where worldsheet time ran around the
long direction of the tube, which is dominated by IR modes.  In
order to study the UV density of states, including the contribution
of long winding strings, we need to analyze the diagram in the
opposite channel.

Consider a rectangular Euclidean worldsheet torus, with the
worldsheet space direction $\sigma_1$ ranging from $0$ to $2\pi
\tilde\tau_2$, and the worldsheet time direction $\sigma_2$ ranging
from $0$ to $2\pi$.  $\tilde\tau_2$ is related to the modular
parameter $\tau_2$ of the previous section by a modular tranform
$\tilde\tau_2=1/\tau_2$. That is, to study the UV behavior
$\tau_2\to 0$, we are interested in the asymptotic behavior as
$\tilde\tau_2\to\infty$ of this diagram.\foot{We can see the basic
effect we need without keeping track of worldsheet angles determined
by $\tau_1$, which we have set to zero for simplicity. As discussed
further below, at least in familiar examples the exponential growth
of the level-matched and non-level-matched spectra are the same.}



Let us start with the case $n=2$, where the simplest version of the
Selberg trace formula computes for each value of $t$ the path integral
we are interested in, in the degeneration limit
$\tilde\tau_2\to\infty$, as a sum over periodic orbits. Up to an
overall constant, the sum over windings and their transverse motions
contribute to $Z_{\rm{UV}}$ as
\eqn\Ztrace{\int dl\rho(l){1\over {\rm
sinh}(l/2t)}e^{-l^2/(4\pi\alpha^\prime\tilde\tau_2)}\qquad (n=2),}
where $\rho(l)$ is the density of periodic geodesics.
The latter is given \refs{\margulis, \manningtop, \bogomolny} by
\eqn\rhola{\rho(l)=e^{l/t} \qquad\qquad (n=2) }
(up to powers of $l$).  Here we are using the fact that $t$ is the
curvature radius of ${\cal M}_n$ at time $t$.

Altogether, working out the saddle point approximation to the $l$
integral in \Ztrace\ at large $l$ (large $\tilde\tau_2$) gives
\eqn\checka{Z_{\rm{UV}}\sim {\rm exp}\left({{\pi\alpha'}\over{4
t^2\tau_2}}+{{\pi c_{\rm
eff}^{\rm{crit}}}\over{6\tau_2}}\right)\qquad\qquad (n=2),}
%
%
%
where $c_{\rm{eff}}^{\rm{crit}}$ arises from the usual sum over
oscillator contributions.  This result \checka\ reproduces directly
in the winding channel the enhancement to the effective central
charge needed to satisfy the modular invariance requirement
\ZUVorb.\foot{Here we are computing the density of states for
physical winding strings, whereas the trace appearing in the
partition function \genform\ includes all states in the CFT, whether
or not they are level matched. In the familiar case of flat space
the density of states at high level scales the same for
level-matched and non-level-matched states (though the subleading
polynomial dependence differs).  Our check will work for the
level-matched states, which leads us to suspect that they scale the
same as the non-level-matched states also in our system.}

The form of \Ztrace\ can be understood in a Hamiltonian description
directly in the winding string channel.  In particular, we interpret
the factor of $1/{\rm sinh}(l/2t)$ in \Ztrace, which is novel relative
to the simpler case of toroidal target spaces, as coming from the sum
over transverse motions of the winding string. Unlike in the case of a
toroidal target space, in our negatively curved target space the
geodesics wrapped by the winding strings are shorter than any other
curves in the same free homotopy class; moving the winding string
in the normal direction, away from the minimal geodesic, will increase
its length and hence its energy. So the transverse degrees of freedom
will produce an approximate harmonic oscillator spectrum.\foot{For
strings wound on a flat cylinder, the analogous sum over transverse
positions generates only a subexponential contribution to the UV limit
of the partition function.}

In the winding channel of our diagram, the worldsheet space
coordinate $\sigma_1$ runs from 0 to $2\pi\tilde\tau_2$, while the
worldsheet Euclidean time direction $\sigma_2$ runs from 0 to
$2\pi$.
Integrating along the $\sigma_1$ (worldsheet space) direction, this
leads to a Hamiltonian for the transverse motions of the form
\eqn\Hamtrans{H= {\alpha^\prime\over 2}\biggl(
p_\perp^2\tilde\tau_2+{l[X_\perp]^2\over{\tilde\tau_2(2\pi\alpha^\prime)^2}}\biggr)+\cdots,}
where the functional dependence $l[X_\perp]^2\sim
l^2(1+X_\perp^2/t^2) $ incorporates the stretching of the string in
the transverse directions in the constant curvature geometry, and
$(\cdots)$ indicates terms independent of the normal degrees of
freedom $X_\perp,p_\perp$.  The frequency of the resulting harmonic
oscillator spectrum is $\omega = l/(2\pi t)$, and the Euclidean time
direction in this channel is $\sigma_2\in (0,2\pi)$.  So we have the
following contribution from the transverse modes:\foot{This
interpretation owes its origin to some observations in \micha, where
a similar factor also arises from a harmonic oscillator spectrum (in
that case from the time variable).}
\eqn\transosc{\sum_{N=0}^\infty e^{-(N+1/2)l/t}={1\over {2{\rm
sinh}(l/2t)}}.}

The generalization to arbitrary $n$ is similar.  The analogue of
\Ztrace\ is now
\eqn\Ztracen{\int dl\rho(l)\biggl({1\over
{\rm
sinh}(l/2t)}\biggl)^{n-1}e^{-l^2/(4\pi\alpha^\prime\tilde\tau_2)}.}
The $n-1$ powers of $1/{\rm sinh}(l/2t)$ arise because there are $n-1$
transverse spatial directions $X_\perp$, each yielding a harmonic
oscillator spectrum as in \transosc.  The density of geodesics is, for
general $n$,
\eqn\rhol{\rho(l)=e^{(n-1)l/t}}
(again up to powers of $l$) \refs{\margulis, \manningtop, \bogomolny}.
The saddle point approximation now gives
%
%
\eqn\checkan{Z_{\rm{UV}}\sim {\rm exp}\left({{\pi\alpha'(n-1)^2}\over{4
t^2\tau_2}}+{{\pi c_{\rm eff}^{\rm{crit}}}\over{6\tau_2}}\right),}
which again exactly reproduces the requirement \ZUVorb.

Thus we see that the sum over the exponential density of winding
modes provides a supercritical contribution to the effective central
charge, related by modular invariance to the IR pseudotachyon. This
test of modular invariance is satisfied by virtue of the
mathematical relation between geometry (producing pseudotachyons)
and topology (producing a Hagedorn spectrum of winding modes)
\refs{\milnor,\margulis,\manning}.  It would be very interesting to
work out the full partition function before doing the $t$ integral
to understand the structure of the full string-theoretic amplitude.
In any case the relation between topology and geometry is evident in
the deep UV and deep IR limits.

%
%

\subsec{Fermionic Contributions in the Superstring}

Let us next address the contribution of fermionic states.
These contribute with opposite sign from the bosons, and {\it a priori} they
may affect the net $c_{\rm eff}$ obtained from the partition
function.  For the following reasons, we expect the contribution to
$c_{\rm eff}$ that we calculated above to survive in the full
partition function, including the fermionic contributions.

First, let us discuss this question in the IR limit.  In the parent
flat spacetime theory, the fermion contribution cancels that of the
bosons.  In the daughter theory, we saw that the bosons develop new
pseudotachyonic instabilities arising from the newly normalizable
$k=0$ modes. We do not expect the fermions to behave in the same
way.  The $k=0$ mode is not generally an eigenstate of the Dirac
operator which respects the spin structure and the reality of the
action.  In general, the hermiticity of the Dirac Hamiltonian
$i\gamma_\mu\nabla^\mu$ would seem to preclude pseudotachyonic
fermions.

Let us discuss this question next in the UV limit. Recall that in
tachyon-free and pseudotachyon-free theories, the high energy
density of states of bosons and of fermions must cancel exquisitely
\kutsei.  In the present case, the exponential density of winding
modes is exactly that required for modular invariance provided it
does not cancel out in the sum over bosonic and fermionic states.
This non-cancellation is a generic expectation which can be
motivated in our case as follows.  Let us consider separately the
sum over winding sectors and the sum over eigenstates of $L_0+\tilde
L_0$ in each winding sector (the latter we will loosely refer to as
the oscillator levels). In the unwound sector, the SUSY breaking is
detected only by paths which propagate around a nontrivial cycle
introduced by the projection to a compact space, since the parent
theory was simply flat spacetime. These contributions are suppressed
by a factor of $e^{-R(t) m}$ for a cycle of length $R(t)$, so
summing over these high oscillator levels weighted by the Hagedorn
density $e^{m\sqrt{\alpha'}\pi\sqrt{2c_{\rm{crit}}/3}}$ does not
yield a divergence in itself.  The same might be expected for fixed
winding number at high oscillator level $m\gg R(t)/l_s^2$, since the
winding is a subleading contribution to the sum over random walks
traced out by the string at high oscillator level (though to check
this would require a more detailed accounting of the way in which
the asymptotic SUSY cancellation of oscillator levels works). The
sum over winding sectors at fixed oscillator level (say the zero
mode) probes the supersymmetry breaking directly, and there is no
reason to expect a similar exponential suppression.

It would be interesting to work out the fermion contributions in
more detail.  An important issue in that regard is the contribution
of the worldsheet fermion determinants to the semiclassical path
integral.  Since the IR propagator for a given bosonic field is the
same in the superstring as it is in the bosonic theory, this
determinant must not change the leading $l$ dependence in the
expression \Ztracen. This seems plausible to us since the dependence
on the winding string length $l$ arises most directly in the
boundary conditions on the bosonic embedding coordinates $X^\mu$,
and its role in the fermionic sector is more indirect. A full
calculation of the fermionic contributions might make use of known
variants of the Selberg trace formula involving superparticles. Such
formulas would need to be used with care, however, to properly take
into account the embedding of the spatial slices into the full
spacetime.

The application of our correspondence to the superstring has the
feature that the critical contribution to $c_{\rm{eff}}$ vanishes, so
that the winding mode contribution is the only source for $c_{\rm{eff}}$.
As we just discussed, we expect this contribution to survive the
supertrace over bosons and fermions.  Given this, the ``ten
dimensional" superstring theory on a (large) finite-volume compact
hyperbolic target space is in fact a (slightly) supercritical
background of string theory.

\subsec{Generalization}

Milnor's theorem \milnor\ applies to arbitrary manifolds of negative
sectional curvature.  The proof uses a result of G\"unther that for
any manifold $\M$ of strictly negative sectional curvature $\le
-\alpha^2$, the volume $V(y)$ of a ball of proper radius $y$ about a
point $y_0$ in its universal cover $\UM$ satisfies
\eqn\Vbound{V(y) \ge
{{n\omega_n}\over{(n-1)(2\alpha)^n}}e^{(n-1)\alpha y}}
for $y\to\infty$.  Here $\omega_n$ is the volume of a unit ball in
$\IR^n$.

In some cases, compact manifolds may be obtained
by quotienting by
isometries in the covering space. It would be interesting to
understand if the behavior \Vbound\ combined with the strictly
negative sectional curvature always entails a gap in the spectrum of
the right value to match the supercritical effective central charge
arising from the winding modes, similarly to what we found for the
constant curvature cases.

However, we should emphasize that a gap in the spectrum
of the covering space is
sufficient but not necessary to obtain pseudotachyons.  In
particular, there are examples which are not of strictly negative
sectional curvature, and have exponential volume growth and hence
exponential growth of the fundamental group, but no gap in the
spectrum.  We will turn to this case in the next subsection to see
how the requisite pseudotachyons arise.

In any case, for consistent string backgrounds on more general
manifolds of negative sectional curvature, Milnor's theorem
guarantees that extra effective dimensions emerge at finite volume.
The fact that negative sectional curvature leads to pseudotachyons
is intuitively plausible. Consider a scalar field $\eta$ propagating
on a compact manifold of negative sectional curvature. The negative
curvature locally causes the space to stretch out rapidly enough
that the friction on spatial modes of $\eta$ can compete with the
gradient energy in the modes.  (If the gradient energy dominated at
late times, then the modes would have positive effective mass
squared; when the friction energy competes at late times this can
lead to pseudotachyons \refs{\ofer,\HellermanNX} in the way we saw
above for the constant curvature case.)

It is clear moreover that the manifolds yielding pseudotachyons in
the infrared include spaces without strictly negative sectional
curvature.  As long as there is a region with negative sectional
curvatures, in which the modes of $\eta$ have support, the friction
will compete with the gradient energy and can produce a
pseudotachyon. Turning this around, if modular invariance holds we
can also conclude that any manifold with exponentially growing
fundamental group must evolve with time in such a way as to contain
regions expanding fast enough to produce a pseudotachyon.

\subsec{Sol Geometry}

An interesting class of examples with exponential growth is obtained
from compact quotients of the Sol geometry by discrete isometries.
In these examples, the fundamental group grows exponentially but the
universal covering space, the Sol geometry
\eqn\Solspace{ds^2=dz^2+{1\over 2}e^{-2z}dx^2+{1\over 2}e^{2z}dy^2,}
has no gap in its spectrum\foot{We thank Michael Freedman for pointing out these
features to us.}. Nonetheless
we will see that there are pseudotachyons
in the spectrum, as needed for modular invariance.

A simple way to obtain solutions for the evolution of homogeneous
but anisotropic spaces is to include many extra dimensions $D\sim
Q^2\gg 1$, in a timelike linear dilaton solution.  The resulting
friction renders the time evolution close to the RG flow of the
corresponding sigma model.    At large radius, the latter in turn
reduces to Ricci flow, for which simple solutions have been worked
out \Riccisolns. Let us adopt this prescription for convenience, to
illustrate the consistency of our methods in the case of the Sol
geometry using the simplicity of the evolution under Ricci
flow.\foot{To match to the usual mathematical conventions for Ricci
flow, we will set $\alpha'=2$ in this subsection and in \S4.2.} In
this setup, the effective central charge is large ($\sim D$) to
begin with, and we will be interested in picking out the enhancement
to it arising from the exponentially growing fundamental group in
the Sol component of the target space.

For the Sol geometry, the late-time solution reviewed for example in
\Riccisolns\ translates into a string-frame metric
\eqn\sollate{ds^2_S=-dt^2+{{4t}\over Q}\,dz^2+{1\over
2}e^{-2z}dx^2+{1\over 2}e^{2z}dy^2+dx_\perp^2,}
where the RG scale is $\mu e^{t/Q}$ and the extra dimensions are
taken to be flat with coordinates $x_\perp$. The string coupling
decreases exponentially with $t$: $g_s\sim g_0 e^{-Qt}$.

Compactifying the $x_\perp$ directions on a flat torus, and
converting to Einstein frame in the remaining four dimensions,
yields a metric
\eqn\Esol{ds^2_{4d,E}=e^{2Qt}\biggl(-dt^2+{{4t}\over Q}dz^2+{1\over
2}e^{-2z}dx^2+{1\over 2}e^{2z}dy^2\biggr).}
The Laplacian for a scalar field $\eta$ on this spacetime is
\eqn\lapsol{\nabla^2\eta=-{1\over
e^{4Qt}\sqrt{t}}\del_t(\sqrt{t}e^{2Qt}\del_t\eta)+e^{-2Qt}\biggl({Q\over{4t}}\del_z^2\eta+2(e^{2z}\del_x^2\eta
+e^{-2z}\del_y^2\eta)\biggr).}
If we look at the modes $\eta_{00}$ with $k_x,k_y=0$, we get an
equation of motion
\eqn\solreduced{e^{2Qt}\nabla^2\eta_{00}=-\ddot\eta_{00}- \left[
{{1\over {2 t}}+2Q}\right]\dot\eta_{00}-{Q k_z^2\over{4t}}\eta_{00}
= 0.}

Relative to the timelike linear dilaton background, a new
contribution to the Hubble friction (the first term in square
brackets) is manifest in this equation. Moreover, this new
contribution to the Hubble friction $H$ from the expanding $z$
direction competes with the gradient term for modes with nonzero
$k_z$, just as in our spectrum \onshell\ arising in the constant
curvature case discussed in \S3.  In general, in an expanding space
the effective mass squared for a canonically normalized scalar has a
contribution proportional to $-H^2$ \ofer\ (as well as a
contribution of order $\dot H$, which in our example is subleading).
If the magnitude of this is greater than the gradient energy at late
times, then this enhances the IR divergence in the one-loop
partition function. More prosaically, a differential equation of the
form \eqn\diffeq{0 =  \ddot \eta - 2H(t) \dot \eta + m_0^2(t) \eta }
may be rewritten (by substituting $ \eta = a(t) y(t)$, with $
H={\dot a/ a } $) as \eqn\oscillator{ 0 = \ddot y + m_{\rm eff}^2 y
} with \eqn\defofmeff{ m_{\rm eff}^2 \equiv m_0^2 - H^2  + \dot H.}
In
the late-time Sol cosmology, $-H^2$ has a contribution going like
$-Q/t$, which is bigger than the gradient term $Q k_z^2/t$ for
sufficiently small $k_z^2$ (and is also bigger than the additional
contribution $\dot H\sim -1/t^2$ to the effective mass).  This
provides the requisite new pseudotachyonic contribution to
$m^2_{\rm{min}}(t)$ in our background. The analysis in \ofer\
applies directly to a field canonically normalized with respect to
the proper time $e^{Qt}/Q$; it is straightforward to check that a
similar scaling argument yields the same conclusion in that
description.

There is no gap in the spectrum of the covering space, which is the
Sol geometry \Solspace.  But there is still an important distinction
between the covering space and the closed manifold we obtain via
projection.  Namely, the modes with strictly zero momentum in the
$x$ and $y$ directions, those with $k_x=k_y=0$, do not exist as
normalizable modes on the covering space.  For any nonzero $k_x$ or
$k_y$, however small, the additional friction term does not compete
with the gradient term at late times, so on the covering space there
are no pseudotachyons beyond those already present in the original
timelike linear dilaton background. On the compact space obtained by
orbifolding by isometries, the $k_x=k_y=0$ mode is normalizable, and
as just discussed this mode {\it is} pseudotachyonic.

\newsec{Polynomial Growth}

Next consider the case of critical string theory on target spaces
with fundamental groups of polynomial growth.  Since in this case
there is not a Hagedorn density of winding strings,
$c_{\rm{eff}}=c_{\rm{crit}}$, and modular invariance requires that the small
perturbation spectrum contain no pseudotachyons.

\subsec{Non-negative Curvature}

In \milnor, Milnor proved that manifolds with non-negative sectional
curvatures have polynomial growth of the fundamental group.  In
particular, the power law growth for an $n$-manifold of this type is
$\rho(l)\le l^n$. Thus such spaces do not have an enhancement of
their effective central charge.

\subsec{Nil Manifold}

The Nil manifold ${\cal N}$ is an interesting case, with constant
negative scalar curvature but without universally negative sectional
curvatures. Although the growth of the fundamental group in this
case is only power law, the number of words of length $s$ grows like
a constant times $s^4$ (see \eg\ \delaharpe\ sections VI.6, VII.7,
VII.22). This is a larger power than for $T^3$, or any 3-manifold of
positive semidefinite curvature, whose fundamental group grows at
most like $s^3$ \refs{\milnor,\delaharpe}.  This comparison suggests
that even for critical $c_{\rm{eff}}$, the number of degrees of
freedom encoded in the subexponential contribution to the density of
states increases with the potential energy (since negative scalar
curvature corresponds to positive potential energy). This might
provide a generalization of \tendim\ to the closed string sector,
with increasing potential energy corresponding to increasing numbers
of degrees of freedom.

Following \Riccisolns, we can consider the late-time Ricci flow in a
very supercritical limit of our system, as in the above discussion
of the Sol geometry in \S3.5, which introduced the relevant
framework and notation. In the Nil case, the string-frame metric is
at large $t$
\eqn\Nilmet{ds^2_S=-dt^2+{Q^{1/3}\over{3 t^{1/3}}} \left(dx+{1\over
2}ydz-{1\over 2}zdy\right)^2+{t^{1/3}\over
Q^{1/3}}(dy^2+dz^2)+dx_\perp^2,}
leading to a $4d$ Einstein frame metric
\eqn\NilE{ds^2_{4d,E}=e^{2Qt}\biggl(-dt^2+{Q^{1/3}\over{3
t^{1/3}}}\left(dx+{1\over 2}ydz-{1\over 2}zdy\right)^2+{t^{1/3}\over
Q^{1/3}}(dy^2+dz^2) \biggr)}
and a scalar Laplacian
%
\eqn\Nillap{\eqalign{ e^{2Qt}\nabla^2\eta= & -\del_t^2\eta- \left(
{1\over {6 t}}+2Q\right) \del_t\eta
+ 3 {t^{1/3}\over Q^{1/3}}\del_x^2\eta \cr
&+{Q^{1/3}\over t^{1/3}}
\left(\del_z^2\eta+\del_y^2\eta+ ( -{y}\del_z+{z} \del_y )
\del_x\eta + {1\over 4} (y^2+z^2) \del_x^2\eta \right) .
%
}}

We are interested in the possibility of an enhanced IR divergence in
the compact quotient of the Nil geometry.  Such new pseudotachyons
would require normalizable modes for which Hubble friction dominates
at late times.  Since the $x$ direction shrinks, modes with momentum
in the $k_x$ direction blueshift, with gradient energy dominating. Let
us therefore consider $k_x=0$ modes.  For these, \Nillap\ boils down
to
\eqn\Nillapzero{-\del_t^2\eta-\left[ {1\over {6
t}}+2Q\right]\del_t\eta+{Q^{1/3}\over t^{1/3}}(\del_z^2\eta +
\del_y^2\eta).}
As in the Sol case, we can read off a new contribution to the Hubble
friction arising from the Nil directions (the first term in square
brackets).
In contrast to the Sol case, here the new contribution to
the Hubble friction squared is subdominant to the gradient energy in
the $y,z$ directions.
This means these modes should not gain a new pseudotachyonic
contribution in addition to the usual one from the supercritical
linear dilaton, which is consistent with the absence of exponential
growth of the fundamental group.  Also as in Sol, one obtains a
similar result for the field canonically normalized with respect to
the proper time $e^{Qt}/Q$.

\newsec{Intermediate Growth?}

There exist groups with {\it intermediate growth}, meaning the
following.  Express each group element minimally as a ``word" formed
from generators of the group.  A group of intermediate growth has
the property that the number of group elements $\rho(s)$ of word
length $\le s$ grows at large $s$ faster than any power of $s$, but
slower than $e^{\lambda s}$ for any constant $\lambda$ \grigreview.
However, all known examples are not finitely presentable. This has
raised the question of whether or not there exist finitely
presentable groups of intermediate growth
\refs{\webquestion,\grigreview, \delaharpe}.

Any finitely presentable group can be realized as the fundamental
group of a smooth manifold of dimension $n\ge 4$ (see \eg\
\anyfund). Those manifolds constituting good initial data for GR can
be analyzed using the techniques developed here and in \mutation.
(The requirement of having good initial data is a very weak
constraint, including as in \S3.6\ and \S4.2\ the rolling dilaton in
string theory.)

Suppose, as a particular example of intermediate growth, that the
density of winding modes on such a manifold $\M_I$ grew like
\eqn\intrho{\rho(l)\sim e^{\lambda l^\gamma}}
for some real numbers $\lambda$ and $\gamma$, with $\gamma < 1$. Consider the
perturbative string one-loop partition function on the spacetime
obtained by time-evolving $\M_I$.  The UV limit of the sum over
winding modes would then take the form
\eqn\UVint{\int dl e^{\lambda l^\gamma} f(l)
e^{-l^2 \tau_2},}
where $f(l)$ is the analogue of the $1/\sinh(l/2)$ factors in
\Ztracen.

Ignoring the $f(l)$ factor for a moment, and performing the $l$
integral by saddle point approximation about the stationary point
$l_*^{2-\gamma} =\gamma\lambda/(2\tau_2)$, yields
a winding contribution to $Z_{{\rm UV}}$ of the form
\eqn\lsaddle{
{\rm \exp}\left[ (2-\gamma)
\left({\lambda \over 2}\right)^{2/(2-\gamma)} \left({\gamma \over
\tau_2}\right)^{\gamma/(2-\gamma)}
 \right]
.}

Assuming that smooth solutions of the long-distance equations of
motion of string theory yield consistent, modular-invariant string
backgrounds,\foot{In another context, this question was addressed
recently in \simeonWalcher, where the authors found modular
invariance to arise nontrivially in a basic class of smooth string
backgrounds.} the same one-loop partition function must have as an IR
limit that goes like\foot{In this section we have assumed
$c_{\rm eff}^{\rm crit} = 0$ for simplicity.}

\eqn\IRint{Z_{\rm{IR}} \sim {\rm
exp}\left(-(\rm{const})\tau_2^{\gamma/ (2-\gamma)}
\right).}

For the putative theory with intermediate growth, $\gamma < 1$ and
this expression \IRint\ is not of the form \IRZ.  In order to
recover an expression consistent with modular invariance, the factor
$f(l)$ in \UVint\ would have to have the property that it completely
cancelled out the leading $e^{l^\gamma}$ behavior of the density of
states.  A similar cancellation would have to take place for any
other intermediate growth function.  While this seems unlikely in
general, one could alternatively take it as a hint for how to
attempt to construct manifolds that exhibit fundamental groups with
intermediate growth.

\newsec{Discussion:  Lessons, Generalizations, Applications}

So far we have seen in prototypical examples how the relation
between geometry and topology fits neatly with perturbative string
modular invariance.  Let us now make some comments about
mathematical and physical implications and generalizations.

\subsec{Physical Lessons and Analogies}

The emergence of new dimensions arising from the growth of the
fundamental group of compact negatively curved spaces is novel, but
in some ways reminiscent of previous phenomena in string theory. Let
us compare and contrast our phenomenon to other familiar effects.

\item{(i)}  In string/M theory duality, a new dimension arises at
strong coupling in type IIA string theory, and in M theory on a
$T^2$, the small volume limit builds up the 10-dimensional type IIB
theory.

\item{(ii)}  String theory on a very small $T^n$ is equivalent to string
theory on a large $T^n$ via T-duality. Perhaps this is the closest
existing analogue of the effect discussed here.  At very small
radius, there are significant indications \refs{\Dproject,
\mutation}\ that the system naturally asymptotes to a large number
of new dimensions which may be geometrical. However, in the torus
case, the effective central charge is independent of the size; the
effective number of dimensions in which strings can oscillate is
fixed.  In our case, $c_{\rm{eff}}$ grows as the space shrinks.

\item{(iii)} In open string theory, the tension of D-branes is related to the dimension of the
open string Hilbert space \tendim.  Here, we have seen that positive
potential energy arising from negative scalar curvature correlates
with extra degrees of freedom.  In the case of negative sectional
curvature, the new degrees of freedom build up an enhanced
$c_{\rm{eff}}$.  In cases like the Nil geometry with negative scalar
curvature but indefinite sectional curvature, there is no
enhancement of $c_{\rm{eff}}$,  but we saw that the power law growth of
the fundamental group exceeded that of any non-negatively curved
target space.

\subsec{Scope}

Milnor's proof \milnor\ implies an exponential density of winding
modes in any space of strictly negative sectional curvature. String
theoretic modular invariance predicts a supercritical $c_{\rm{eff}}$
for any consistent perturbative background with pseudotachyons in
its spectrum of small fluctuations.  We have seen that these two
phenomena line up precisely in the examples discussed above.  In
more general circumstances, one or the other notion may apply
independently.

It is interesting to ask whether the correspondence exhibited here
extends to perturbatively metastabilized solutions, such as
\newhandle.  In particular, de Sitter and inflationary solutions
have pseudotachyonic modes.  To the extent that perturbative string
theory applies, which in metastabilized solutions would require
applying the Fischler-Susskind method for retaining conformal
invariance \FS,  these modes translate via modular invariance to an
enhanced $c_{\rm{eff}}$.  This interpretation is consistent with the
present state of the art in moduli stabilization, which exhibits the
following pattern. Low energy supersymmetric de Sitter models have
crucial contributions from non-perturbative ingredients \kklt, so
string perturbation theory (even at the level of Fischler-Susskind)
does not apply and $c_{\rm{eff}}$ is not well defined. Perturbative
models with SUSY broken at or above the Kaluza-Klein scale in the
geometry are explicitly supercritical \mss\ or involve a negatively
curved compactification
\newhandle, which we now also understand to be supercritical.

\subsec{Future Directions and Applications}

There are many directions that suggest themselves. First, it is
worth emphasizing that in the case we studied here, we analyzed only
the deep UV and deep IR physics of the one-loop amplitude, focusing
on the leading bosonic contributions. It would be very interesting
to work out the fermionic contributions in detail, and to compute
(if possible at finite time) the full one loop partition function
including all the ``oscillator" modes.

Mathematically, the relation between geometry and topology is a
subject of current research, for example as discussed in the above section on
intermediate growth. It arises in the classification of 3-manifolds
(which in effect constitute $(3+c_{\rm{eff}})$-dimensional spaces in string
theory, suggesting perhaps an alternate classification scheme). For
another example, the full characterization of groups that can appear
as fundamental groups of K\"ahler manifolds is an open problem. It
is interesting to consider whether a more detailed probe than the
one-loop partition function could help determine the limitations on
fundamental groups of K\"ahler manifolds \tomasiello.

We focused here on the compact case, for which Milnor's theorem
\milnor\ is sufficient to derive the new effective dimensions from
winding strings.  Compactness is a sufficient, but not necessary,
condition; we expect there to be examples of {\it localized}
pseudotachyons which correspond to an exponential density of winding
modes arising in {\it noncompact} manifolds with exponentially
growing fundamental group.  It might be interesting to study these
cases in detail.

At large radius, it is interesting to ask whether group theoretic
considerations can help determine features of the evolution and
small perturbation spectrum of these spaces.  For example, it is of
interest to learn what cosmological phases arise from vacuum
solutions obtained from compactification on generic curved
manifolds.  Initial conditions in such string cosmologies will be
affected by the behavior at small radius, at very early times.

Physically, a key question is indeed the behavior of the system at
very small radius.  There are several independent indications that
the system can naturally asymptote to a specific supercritical
background of string theory with a rolling closed string tachyon;
these are under investigation \Dproject.\foot{See \othermutation\
for potentially related recent works.} More specifically, for $n=2$
the Jacobian torus may arise naturally (and for higher even
dimensions the analogous Albanese variety). This is the moduli space
of a single wrapped D-brane probe; moreover for a spherical shell of
wrapped D-branes this torus appears as its approximate moduli space,
within a UV complete system providing a field theory dual. In the
closed string sector alone, in even dimensions, the embedding of the
compact manifold ${\cal M}_n$ into a torus of any dimension proceeds
through the Jacobian/Albanese variety, which preserves the
symmetries of the system by mapping one-cycles of ${\cal M}_{n}$
into those of the torus. Moreover, a preliminary analysis indicates
that a variant of Buscher's path integral derivation of T-duality
produces in our case the Jacobian perturbed by a rolling tachyon.
This substantiates the hint already available at large radius, that
the high energy winding strings trace out a lattice random walk in
$2h$ dimensions for a Riemann surface of genus $h$, suggesting that
a continuum limit of $2h$ dimensions might arise \mutation.

Among homogeneous spaces, negative curvature is generic in
dimensions $n=2$ and $3$. Three being the number of large spatial
dimensions in the observed universe, there is strong motivation for
getting a handle on these spaces in string theory. The winding
string degrees of freedom (which go beyond general relativity) build
up new effective dimensions, and these degrees of freedom become
more important as the space shrinks.\foot{It would be interesting to
understand how this affects the question of summing over
``three-manifolds" (which has interesting subtleties to do with the
distribution of volumes at fixed curvature \threesum).} In more
formal terms, it seems that a new type of string duality (perhaps
best described as a ``D(imensional) duality") is emerging in the
case of negatively curved compact target spaces.

\bigskip
\bigskip
\noindent{\bf Acknowledgements}

\nobreak
We are grateful to A. Maloney for originally pointing out that the
Selberg trace formula provides a precise method for calculating the
enhancement to $c_{\rm{eff}}$, and to O. Aharony for many
discussions on pseudotachyons.  We are grateful to D. Green, A.
Lawrence and D. Morrison for their many insights in the related work
\Dproject. We thank A. Adams, M. Freedman, G. Horowitz, S. Kachru,
N. Lambert, L. McAllister, J. Polchinski, and A. Tomasiello for
useful discussions on this topic. E.S. and D.S. thank the KITP for
hospitality during the completion of this work, and gratefully
acknowledge useful discussions in the UCSB math/physics seminar on
``Geometry, Topology, and Physics" run by D. Morrison. This research
was supported in part by the National Science Foundation under Grant
No. PHY99-07949. E.S. and D.S. are also supported in part by the DOE
under contract DE-AC03-76SF00515, by the NSF under contract 9870115,
and by an FQXI grant.  The work of J.M. is supported in part by
funds provided by the U.S. Department of Energy (D.O.E.) under
cooperative research agreement DE-FG0205ER41360.

\listrefs

\end